\documentclass[useAMS,usenatbib]{mn2e}

\usepackage[latin1]{inputenc}  
\usepackage{graphicx}
\usepackage{txfonts}

\title[Production of H$_3^+$ via molecular photodissociation.]{Production of H$_3^+$ via photodissociation of organic molecules in interstellar clouds}
\author[S. Pilling et al.]{S. Pilling$^{1}$\thanks{E-mail:
spilling@lnls.br}, D. P. P. Andrade$^{2,3}$, R. Neves$^{2}$, A. M.
Ferreira-Rodrigues$^{2,3}$, \and A. C. F.
Santos$^{4}$, H. M. Boechat-Roberty$^{2}$  \\
\\
$^{1}$Laboratório Nacional de Luz Síncrotron, Caixa Postal 6192, CEP
13084-971, Campinas, SP, Brazil.\\
$^{2}$Observatório do Valongo, Universidade Federal do Rio de Janeiro - UFRJ, Ladeira
Pedro Antônio, 43, CEP 20080-090, Rio de
Janeiro, RJ, Brazil.\\
$^{3}$Instituto de Química, Universidade Federal do Rio de Janeiro -
UFRJ, Ilha do Fundão, CEP 21949-900, Rio de Janeiro, RJ, Brazil.\\
$^{4}$Instituto de Física, Universidade Federal do Rio de Janeiro -
UFRJ, Ilha do Fundão, Caixa Postal 68528, CEP 21941-972, Rio de
Janeiro, RJ, Brazil.\\}
\begin{document}

\date{Received / Accepted}

\pagerange{\pageref{firstpage}--\pageref{lastpage}} \pubyear{2005}

\maketitle

\label{firstpage}


\begin{abstract} 
We present experimental results obtained from photoionization and photodissociation
processes of abundant interstellar CH$_3$-X type organic molecules like methanol
(CH$_3$OH), methylamine (CH$_3$NH$_2$) and acetonitrile (CH$_3$CN) as alternative route
for the production of H$_3^+$ in interstellar and star forming environments. The
measurements were taken at the Brazilian Synchrotron Light Laboratory (LNLS), employing
soft X-ray photons with energies between 200 and 310 eV and time of flight mass
spectrometry. Mass spectra were obtained using the photoelectron-photoion coincidence
techniques. Absolute averaged cross sections for H$_3^+$ production by soft X-rays were
determined. We have found that, among the channels leading to molecular dissociation,
the H$_3^+$ yield could reach values up to 0.7\% for single photoionization process and
up to 4\% for process involving double photoionization. The H$_3^+$ photoproduction
cross section due to the dissociation of the studied organic molecules by photons over
the C1s edge (200-310 eV) were about 0.2-1.4 $\times$ 10$^{-18}$ cm$^2$. Adopting the
typical X-ray luminosity $L_X \gtrsim 10^{31}$ erg s$^{-1}$ which best fit the
observational data for AFGL 2591 (Stäuber et al. 2005) we derive an estimative for the
H$_3^+$ photoproduction rate due to methyl-compounds dissociation process. The highest
value for the H$_3^+$ column density from methanol dissociation by soft X-rays, assuming
a steady state scenario, was about $10^{11}$ cm$^2$, which gives the fraction of the
photoproduced H$_3^+$ of about 0.05\%, as in the case of dense molecular cloud AFGL
2591. Despite the extreme small value, this represent a new and alternative source of
H$_3^+$ into dense molecular clouds and it is not been considered as yet in interstellar
chemistry models.
\end{abstract}


\section{Introduction}

The H$_3^+$ ion plays an important role in diverse fields from chemistry to astronomy
such as, the chains of reaction that lead to the production of many of complex molecular
species observed in the interstellar medium (Hersbt \& Klemperer 1973; Dalgarno \& Black
1996; Suzuki 1979). The H$_3^+$ ion was discovered in molecular clouds (McCall et al.
1999; Geballe \& Oka 1996) and in the diffuse interstellar medium (Oka et al. 2005;
McCall et al. 1998) with column densities on the order of $\sim$10$^{14}$ cm$^{-2}$.

As point out by McCall \& Oka (2000), observation of H$_3^+$ can be combined with those
of other important molecules such as H$_2$ and CO to characterize the physical and
chemical properties of the interstellar clouds. Moreover, the physics and chemistry from
H$_3^+$ analysis, combined with low density and temperature of interstellar space, lead
to interesting phenomena like the extraordinary deuterium fractionation, the bistability
of chemical models and the radiative thermalization through forbidden rotational
transitions. A detailed review about the H$_3^+$, the simplest stable interstellar
polyatomic molecule, can be found in Oka (2006).

The main pathway formation of H$_3^+$ occurs via ionization of H$_2$ to H$_2^+$ by the
ubiquitous cosmic ray or local X-ray, followed by the efficient Langevin reaction (Bowers
et al. 1969):
\begin{equation}
 H_2^+ + H_2 \longrightarrow H_3^+ + H
\end{equation}

The main destruction pathway of H$_3^+$ occurs via proton-hop reaction:
\begin{equation}
 H_3^+ + X \longrightarrow HX^+ + H_2
\end{equation}
where X = CO, N$_2$, H$_2$O, NH$_3$, etc. These reaction have been studied by Burt et
al. (1970) and all show Langevin-type behavior. As discussed by Oka (2004), the averaged
H$_3^+$ number density in dense clouds is about 10$^{-4}$ cm$^{-3}$ where X = CO is the
main destroyer. In diffuse clouds, X = $e^-$ is the main destroyer and the averaged
H$_3^+$ number density for these regions is about 10$^{-6}$ cm$^{-3}$. Despite the low
rate coefficients, other formation reaction involving species like HeH$^+$, NH$^+$, HCO
and CH$_3^+$ have also been proposed (McCall et al. 1999; Millar et al.
1997\footnote{UMIST gas-phase chemical reaction network (www.ufda.net)}).

Miller et al. (1992) have observed the H$_3^+$ emission in the infrared spectrum of the
remnants of Supernova 1987A. Brittain \& Rettig (2002) reported a detection of the
H$_3^+$ emission from the Herbig AeBe star HD141569 and speculated that it is from a
protoplanet in the preplanetary disk, however this detection remains controversial (Oka
2002). The H$_3^+$ emission has been found recently in other Herbig Be star LkH$\alpha$
101 by Brittain et al. (2004) with a column density of about $2.2 \times 10^{14}$
cm$^{-2}$.

The observed ubiquity of H$_3^+$ in interstellar clouds suggests that it is also
observable in many other objects where molecules and ionization abound. The intense 3.7
$\mu$m H$_3^+$ emission spectrum from Jupiter (Drossart et al. 1989; Oka \& Geballe
1990; Connerney et al. 1993), Saturn (Geballe et al. 1993) and Uranus (Trafton et al.
1993) has become a general tool to study planetary ionospheres. It was even suggested
that this emission might be detectable from Jupiter-like planets orbiting other stars
(Connerney \& Satoh 2000).

Planetary nebulae and proto-planetary nebulae are also interesting targets. As pointed
by Oka (2004), due to the low metallicity, Magellanic clouds may also be interesting
objects to try H$_3^+$.

The ionized triatomic hydrogen molecule is found as a primary fragment in the
photoionization and electron impact mass spectra of several small molecules. Ruhl et al.
(1990), have reported one of the first studies on the production of H$_3^+$ from methyl
compounds, via two-body charge separation dissociation process due to the absorption of
40.8 eV photons from a toroidal grating monocromator. The authors have pointed out that
the formation of double ionized methyl compound molecules requires excitation energies
of the order of 30 eV. Photon energies of this order of magnitude are not normally
available within dense clouds because absorption by the abundant hydrogen species (and
possibly by interstellar grains) in the other layers screens the hard photons from
penetrating deep into the cloud. However, it is recognized that stars can be buried
within dense clouds and can act as local high energy sources (Charnley et al. 1988)
promoting increase in chemistry complexity.

The H$_3^+$ peak, appearing at m/q = 3 in mass spectra, has also been observed in the
dissociation of other methyl compound molecules, due to VUV photons and electrons, as
the case of C$_2$H$_6$, CH$_3$CN, CH$_3$NCO, CH$_3$COOH, HCOOCH$_3$, CH$_3$NH$_2$,
cyclo-C$_3$H$_6$, n-C$_3$H$_8$, CH$_3$Cl, CH$_3$Br, CH$_3$I and others (Eland 1996;
Thissen et al. 1994). The rates for rearrangement leading to H$_3^+$, D$_3^+$ and
HD$_2^+$ have been studied by Furukawa et al. (2005) in an experiment on methanol and
deuterated methanol (CD$_3$OH) under intense laser (800 nm) fields.

Recently, Sharma \& Bapat (2006) have reported the production of H$_3^+$ ion by the
ionization/dissociation of ethanol molecule by a 1.3 keV electron beam. The authors have
pointed out the determination of active sites for H atom rearrangement in the dissociation
of single and double ionized ethanol (CH$_3$CH$_2$OH) and deuterated ethanol
(CH$3$CH$_2$OD). They also have found that H$_3^+$ is far more likely to be formed by of
rearrangements of H atoms on the CH$_2$ and OH sites, rather than the CH$_3$ site.

In this work, we present a set of experimental results obtained from photoionization and
photodissociation processes of abundant CH$_3$-X type interstellar organic molecules like,
methanol (CH$_3$OH), methylamine (CH$_3$NH$_2$) and acetonitrile (CH$_3$CN), as another
route for the production of H$_3^+$ ion in star-forming regions. In these environments the
radiation field (UV and X-rays) can promote several photophysical and photochemical
processes onto molecules, including the photodissociation. The products of organic
molecules dissociation (ex. reactive ions and radicals) can also provide the formation of
interstellar complex molecules like long carbon chain molecules and amino acids (ex.
glycine).

In section 2 we present the experimental setup and the data analysis technique employed.
The results are shown and described in section 3. Final remarks and conclusions are
present in section 4.

\section{Experimental}

The measurements were taken at the Brazilian Synchrotron Light Laboratory (LNLS), in
Campinas, São Paulo, Brazil. Briefly, soft X-rays photons (100-310 eV) from a toroidal
grating monochromator (TGM) beamline ($\sim10^{12}$ photons/s), perpendicularly intersect
the gas sample inside a high vacuum chamber. The gas needle was kept at ground potential.
The emergent photon beam flux was recorded by a light sensitive diode. The complete
description of the experimental setup could be found elsewhere (Boechat-Roberty et al.
2005; Pilling et al. 2006)

\begin{figure}
\resizebox{\hsize}{!}{\includegraphics{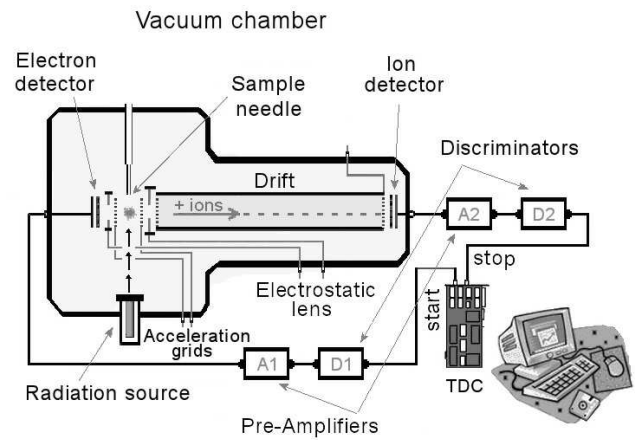}}
\caption{Schematic diagram of the time of flight mass spectrometer inside the experimental
vacuum chamber and the associated electronics.} \label{fig-diagram}
\end{figure}

Conventional time-of-flight mass spectra (TOF-MS) were obtained using the photoelectron
and photoion coincidence (PEPICO) technique. The ionized recoil fragments produced by the
interaction with the photon beam were accelerated by a two-stage electric field and
detected by two micro-channel plate detectors in a chevron configuration, after
mass-to-charge (m/q) analysis by a time-of-flight mass spectrometer (297 mm long). They
produced up to three stop signals to a time-to-digital converter (TDC) started by the
signal from one of the electrons accelerated in the opposite direction and recorded
without energy analysis by two micro-channel plate detectors. A schematic diagram of the
time of flight spectrometer, mounted inside the experimental vacuum chamber, is shown in
Fig.~\ref{fig-diagram}, where A1 and A2 are the pre-amplifiers and D1 and D2 are the
discriminators. The connection to the time-to-digital converter is also shown.

Besides PEPICO spectra, other two kinds of coincidence mass spectra were obtained
simultaneously, PE2PICO spectra (photoelectron photoion photoion coincidence) and
PE3PICO spectra (photoelectron photoion photoion photoion coincidence) (see details in
Pilling 2006b; 2006c). These spectra have ions coming from double and triple ionization
processes respectively, that arrive coincidentally with photoelectrons. Of all signals
received by the detectors only about 10\% come from PE2PICO and 1\% from PE3PICO
spectra, due to the limited detection efficiencies. This reflect that the majority
contribution to data came from aborted double and triple coincidence events.
Nonetheless, PEPICO, PE2PICO and PE3PICO signals were taken into account for
normalization purposes. Negative ions may also be produced and detected, but the
corresponding cross-sections are negligible.

The samples were bought commercially from Sigma-Aldrich with purity greater than 99.5\%.
No further purification was performed other than degassing the liquid sample by multiple
freeze-pump-thaw cycles before admitting the vapor into the chamber.

The base pressure in the vacuum chamber was in the $10^{-8}$ Torr range. During the
experiment the chamber pressure was maintained below $10^{-5}$ Torr. The pressure at the
interaction region (volume defined by the gas beam and the photon beam intersection) was
estimated to be $\sim$ 0.1-1 Torr (10$^{15}$-10$^{16}$ molecules cm$^{-3}$). The
measurements were made at room temperature.

The partial ion yield for H$_3^+$ (PIY$_{H_3^+}$) or relative intensities of H$_3^+$
produced due to the photodissociation of the studied organic molecules, was determined
directly from PEPICO spectra by the expression
\begin{equation}
PIY_{H_3^+} = \left( \frac{A_{H_3^+} }{A^{+}_t} \pm \frac{ \sqrt{A_{H_3^+}}+ A_{H_3^+}
\times ER/100}{A^{+}_t} \right) \times 100\%
\end{equation}
where $A_{H_3^+}$ is the area of a H$_3^+$ peak, the $A^{+}_t$ is the total area of the
PEPICO spectrum. The $ER~=~2-4$ \% is the estimated error factor due to the data
acquisition and data treatment.

In a similar manner, for the PE2PICO spectra, we have determined the partial double
coincidence yield for H$_3^+$ (PDCY$_{H_3^+}$), or the relative production of H$_3^+$ in
coincidence with another ion $i$ from the dissociation of double ionized organic molecule,
by
\begin{equation}
PDCY_{H_3^+} = \left( \frac{A_{i,H_3^+} }{A^{2+}_t} \pm \frac{ \sqrt{A_{i,H_3^+} }+
A_{i,H_3^+} \times ER/100}{A^{2+}_t} \right) \times 100\%
\end{equation}
where $A_{i,H_3^+}$ is the number of events in double coincidence of a given ion $i$ and
H$_3^+$ pair and $A^{2+}_t$ is the total number of count of PE2PICO spectra.

The data treatment and analysis of PE2PICO spectra as well the determination of PDCY
were performed using the program described elsewhere (Pilling 2006).

The complete results about the photoionization and photodissociation of methanol by soft
X-rays could be found elsewhere (Pilling et al. 2006c; 2006d). The same experimental study
for the other abundant CH$_3$-X type interstellar organic molecules like methylamine and
acetonitrile are the subject of future publication.

\section{Results and discussion}

\begin{figure}
 \centering
 \resizebox{\hsize}{!}{\includegraphics{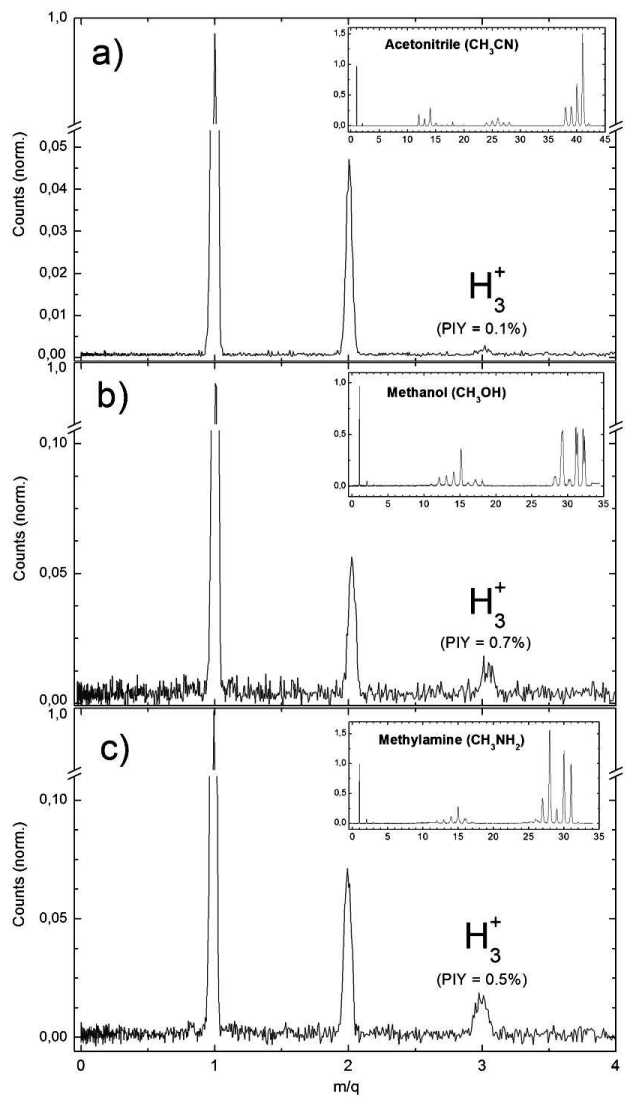}}
\caption{Time of flight mass spectra (PEPICO) obtained by the dissociation of single
photoionized molecules: a) acetonitrile, b) methanol c) methylamine, with the respective
whole spectrum inserted in each figure. See details in text}
 \label{fig:PEPICO-samples}
\end{figure}

Figure~\ref{fig:PEPICO-samples} shows a detail of the mass spectrum of the fragments
produced by single photoionization (PEPICO) of acetonitrile (Fig. 2a), methanol (Fig.
2b) and methylamine (Fig. 2c) recorded with photons at energies between 200 eV and 310
eV (over the C1s resonance, about $\sim$ 290 eV). The presence of the H$_3^+$ peak
(m/q=3) and its relative intensity (PIY) are indicated in each figure. The other two
lightest ions (H$^+$, H$_2^+$) are also seen. The insets in both figures represent the
fully mass spectra of each molecule for comparison. Our data have shown that about
0.1\%, 0.7\% and 0.5\% of the photodissociation channels of acetonitrile, methanol and
methylamine, respectively, lead to the H$_3^+$ production.

\begin{figure}
 \centering
 \resizebox{\hsize}{!}{\includegraphics{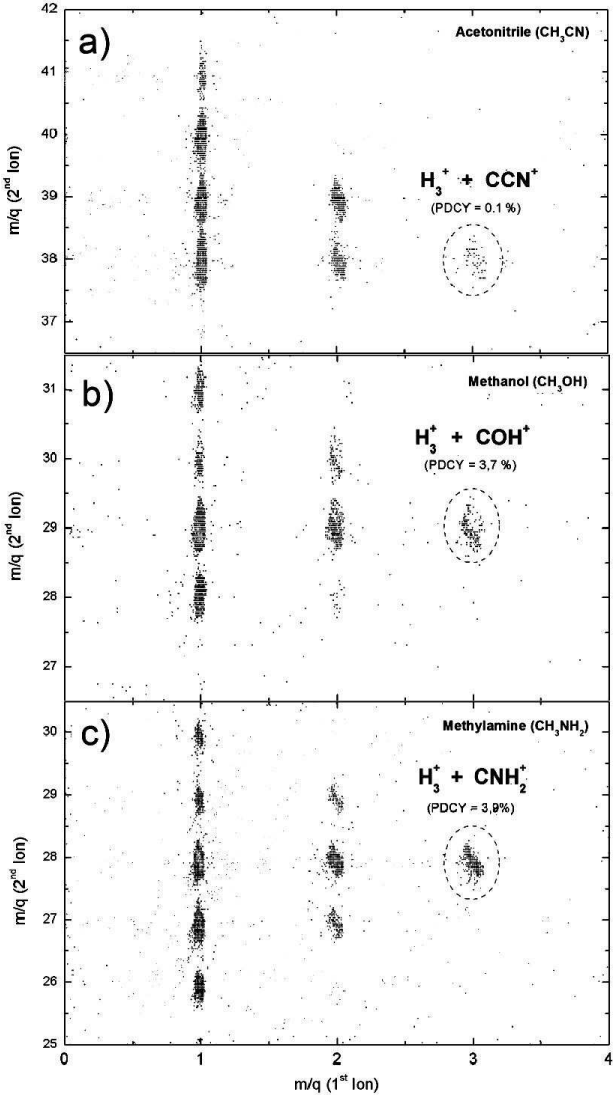}}
\caption{Detail of the PE2PICO spectra obtained by the dissociation of double photoionized
molecules a) acetonitrile, b) methanol c) methylamine. See details in text}
 \label{fig:PE2PICO-samples}
\end{figure}

In star-forming regions like Sgr B2, Orion KL and W51, the presence of widespread UV and
X-ray fields could trigger the formation of photodissociation regions - PDRs (Tielens \&
Hollenback 1985) and X-rays dominated regions - XDRs (Maloney, Hollenbach \& Tielens
1996) where many molecules could be detected (ex. Ehrenfreund \& Charnley 2000). As
pointed out by Casanova et al. (1995), Koyama et al. (1996) and Imanishi et al. (2001),
protostars are extremely efficient sources of X-ray photons that are capable of
traversing large column densities of gas before being absorbed. The X-ray-dominated
regions (XDRs) in the interface between the ionized gas and the self-shielded neutral
layers could influence the selective heating of the molecular gas. The complexity of
these regions possibly allows a combination of different scenarios and excitation
mechanisms to coexist (Goicoechea et al. 2004).

In our previous works (Pilling 2006b; 2006c) we reported that the X-rays photons also
produce double and triple ionization process (ex. Auger process) and, depending on the
photon energy it will be the most effective ionization process. In
Fig.~\ref{fig:PE2PICO-samples}a, b and c, we present a detail of the averaged PE2PICO
mass spectra obtained with photons energies around C1s edge, in the energy range of 200
to 310 eV. These spectra have present several different ions pair coming from different
dissociation channels. The dissociation channels that involves the H$_3^+$ production
and it relative intensity on the PE2PICO spectra (PDCY) are indicated. Our data have
shown that the H$_3^+$ production represents about 0.1\%, 3.7\% and 3.9\% of the double
photodissociation channels of acetonitrile, methanol and methylamine, respectively. A
comparison between the PDCY$_{H_3^+}$ recorded with VUV photons (Eland 1996) with soft
X-ray photons from this work is present in Table~\ref{tab-Comparison}. Despite the small
values on the H$_3^+$ photoproduction by soft X-rays compared with the VUV photons,
inside molecular clouds the soft X-ray field could be higher than the VUV field.

\begin{table}
\centering \caption{Comparison between the partial double coincidence yield for H$_3^+$
ion (PDCY$_{H_3^+}$) recorded with VUV and soft X-ray photons.} \label{tab-Comparison}
 \setlength{\tabcolsep}{4pt}
\begin{tabular}{ l c c  }
\hline \hline
CH$_3$-X molecule     & VUV$^a$     & Soft X-rays$^b$    \\
                      & (40.8 ev)   & (200-310 eV)   \\
\hline
Acetonitrile          & 1.0         & 0.1 / 0.1$^c$     \\
Methanol              & 11.2        & 3.7 / 0.7$^c$    \\
Methylamine           & 13.0        & 3.9 / 0.5$^c$    \\
\hline
\multicolumn{3}{l}{$^a$ Eland (1996); $^b$ this work;  }\\
\multicolumn{3}{l}{$^c$ PIY (relatives intensities from single ionization)  }\\
\end{tabular}
\end{table}

Thissen (1993) have also reported the presence of H$_3^+$ (and also D$_3^+$, HD$_2^+$
and H$_2$D$^+$) on PE2PICO spectra of methylamine and deuterated-methylamine species
recorded by 30-60 eV photons. Following the author, the production of H$_3^+$-like
species from dissociation of single ionized methylamine-like molecules at these photon
energies or lower is negligible.

\subsection{H$_3^+$ photoproduction cross section and rate in soft X-ray field}

As pointed by Stäuber et al. (2005), due to the low absorption cross section, X-rays can
penetrate deeper into the molecular clouds than, for example, UV photons and affect the
gas-phase chemistry even at large distance from the source.

The averaged cross sections for H$_3^+$ production by soft X-rays photons in the energy
range of 200 to 310 eV, were determined using the methodology describe elsewhere
(Boechat-Roberty et al. 2005; Pilling et al. 2006a). Following the discussion of Chen et
al. (1981) about the negligible fluorescence yield (due to the low carbon atomic number)
and anionic fragments production in the present photon energy range, we adopted that all
absorbed photon leads to cationic ionizing process. Therefore, in order to put our data
on an absolute scale, after a subtraction of a linear background and false coincidences
coming from aborted double and triple ionization, we have summed up all the
contributions of all cationic fragments detected and normalized them to the absolute
photoabsorption cross sections taken from literature. For acetonitrile and methanol, the
absolute photoabsorption cross section were obtained by Hitchcock et al. (1989) and by
Ishii \& Hitchcook (1988), respectively. In the case of methylamine, due to the lack of
the absolute photoabsorption measurements on literature, we have used the values
obtained for the methanol (Ishii \& Hitchcook 1988) to gives an estimative of the
averaged cross sections for H$_3^+$ production.

Briefly, the photoproduction cross section of H$_3^+$ by the dissociation of single
ionized, $\sigma^+_{H_3^+}$, and from double ionized, $\sigma^{++}_{H_3^+}$, organic
molecules are given by
\begin{equation}
\sigma^+_{H_3^+} = \sigma^{+} \frac{PIY_{H_3^+}}{100} \quad \textrm{and} \quad
\sigma^{++}_{H_3^+} = \sigma^{++} \frac{PDCY_{H_3^+}}{100}
\end{equation}
where $\sigma^{+}$ and $\sigma^{++}$ are the cross section for single ionization and
double ionization of parental organic molecule, respectively (see Boechat-Roberty et al.
2005; Pilling et al. 2006). The $PIY_{H_3^+}$ and $PDCY_{H_3^+}$ are the relative H$_3^+$
yield from PEPICO and PE2PICO spectra as previously discussed. The estimated experimental
error is considered to be lower than 30\%.

\begin{table*}
\centering \caption{Averaged H$_3^+$ photoproduction cross section and photoproduction
rate for an X-ray luminosity of $L_x \gtrsim 10^{31}$ erg/s (Stäuber et al. 2005), from
the dissociation of methanol, methylamine and acetonitrile by soft X-rays photons over
the C1s edge (200-310 eV). See details in text.}
 \label{tab-sigma}
 \setlength{\tabcolsep}{5pt}
\begin{tabular}{ l c c c c}
\hline
CH$_3$-X molecule  & $\sigma^+_{H_3^+}$             & $\sigma^{++}_{H_3^+}$            &  $\sigma_{H_3^+} = \sigma^+_{H_3^+}$ + $\sigma^{++}_{H_3^+}$ &  k$_{ph}$\\
                   & ( $\times 10^{-19}$ cm$^{2}$)  &   ( $\times 10^{-19}$ cm$^{2}$)  &  ( $\times 10^{-18}$ cm$^{2}$)  &  ( $\times 10^{-15}$ s$^{-1}$) \\
\hline
Acetonitrile          & 2.0                         & 0.2                              & $\sim$ 0.2           & $\gtrsim$ 40$^b$;    $\gtrsim$ 0.05$^c$    \\
Methanol              & 12.0                        & 2.0                              & 1.4                  & $\gtrsim$ 300$^b$;   $\gtrsim$ 0.4$^c$ \\
Methylamine$^a$       & 8.0                         & 3.0                              & 1.1                  & $\gtrsim$ 200$^b$;   $\gtrsim$ 0.3$^c$ \\
\hline
\multicolumn{5}{l}{$^a$ Estimated value.}\\
\multicolumn{5}{l}{$^b$ At a distance $r \sim$ 200 AU ($2.5 \times 10^{15}$ cm) from the central source; $F_{softX} \gtrsim 2 \times 10^5 $ photons cm$^{-2}$ s$^{-1}$.}\\
\multicolumn{5}{l}{$^c$ $r \sim$ 5000 AU ($7 \times 10^{16}$ cm); $F_{softX} \gtrsim 3 \times 10^2 $ photons cm$^{-2}$ s$^{-1}$.}\\
\end{tabular}
\end{table*}

In general, as discussed by Stäuber et al. (2005), primary X-ray ionization plays only a
minor role in the chemistry since reaction are $\sim 1000$ times slower compared to the
relevant chemical reactions and more than 10 times slower than electron impact
ionization as the case of AFGL 2591 model parameters. However, in denser regions inside
the molecular clouds, the X-ray ionization rate may exceed the cosmic-ray ionization and
become a significative source of photoionization and photodissociation.

The $H_3^+$ photoproduction rate due to the dissociation of methyl-compound molecules by
soft X-rays (200-310 eV) is given by the simple expression:
\begin{equation}\label{eq-k}
k_{ph} = \int \sigma_{H_3^+}(\varepsilon) F(\varepsilon) d{\varepsilon} \sim
\sigma_{H_3^+} F_{softX} \quad \textrm{[s$^{-1}$]}
\end{equation}
where $\sigma_{H_3^+} = \sigma^+_{H_3^+}$ + $\sigma^{++}_{H_3^+}$ and $F_{softX}$ is the
averaged $H_3^+$ photoproduction cross section and photon flux over the soft X-ray
energy (200-310 eV).

Following Stäuber et al. (2005), the radiative flux in the X-ray domain inside the dense
molecular cloud AFGL 2591 can be described by a black body with temperature $T_{X}
\gtrsim 3 \times 10^{7}$ K resulting a typical X-ray luminosity $L_X \gtrsim 10^{31}$
erg S$^{-1}$ with $L_X/L_{bol} \sim 10^{-6}$. Taking into an account only the thermal
emission, we calculate the soft X-ray photon flux at two representative distances from
de source inside the cloud. At about a distance of $r \sim$ 200 AU ($2.5 \times 10^{15}$
cm), the lower distance used in the AFGL 2591 X-ray chemistry models by Stäuber et al.
(2005). The other adopted distance was at $r \sim$ 5000 AU ($7 \times 10^{16}$ cm) from
the central source, which represents the distance were the H$_2$ ionization rate due to
X-rays is in the same order of magnitude of the constant cosmic-ray ionization rate
($k_{cr} \sim 3-6 \times 10^{-17}$ s$^{-1}$) inside this cloud. The photon flux in the
soft X-ray range (200-310 eV) for these two regions were about $F_{softX} \gtrsim 2
\times 10^5 $ and $\gtrsim 3 \times 10^2$ photons cm$^{-2}$ s$^{-1}$, respectively.

The averaged H$_3^+$ photoproduction cross sections due to single ionization,
$\sigma^+_{H_3^+}$, and due to double ionization, $\sigma^{++}_{H_3^+}$, for the studied
organic molecules, as well the fully H$_3^+$ photoproduction cross section,
$\sigma_{H_3^+}$ and the H$_3^+$ photoproduction rate $k_{ph}$ in the energy range from
200 to 310 eV are shown in Table~\ref{tab-sigma}.

\subsection{Photoproduced H$_3^+$ abundance}

\begin{table*}
\centering \caption{The H$_3^+$ column density due to the photodissociation of methanol
by soft X-rays, $N^{ph}_{H_3^+}$, in some dense molecular clouds. The observed column
density of CH$_3$OH and H$_3^+$ in each region are also given. The last column
represents the fraction of the produced H$_3^+$ due to CH$_3$OH photodissociation,
$N^{ph}_{H_3^+}/N_{H_3^+}$. See details in text.} \label{tab-N}
\begin{tabular}{ l l l l l }
\hline
                         &  $N_{CH_3OH}^a$           & $N_{H_3^+}^b$                     & $N^{ph}_{H_3^+}$                         & $N^{ph}_{H_3^+}/N_{H_3^+}$  \\
Dense Molecular Clouds   & ( $\times 10^{15}$ cm$^{-2}$) & ( $\times 10^{14}$ cm$^{-2}$) & ( $\times 10^{8}$ cm$^{-2}$)             &  ( $\times 10^{-5}$ )                         \\
\hline
AFGL 2136                &  0.44                      & 3.8                              & $\gtrsim$ 400$^c$; $\gtrsim$ 0.7$^d$     & $\gtrsim$ 10$^c$; $\gtrsim$ 0.02$^d$                 \\
AFGL 490                 &  0.36                      & 1.1                              & $\gtrsim$ 400$^c$; $\gtrsim$ 0.5$^d$     & $\gtrsim$ 30$^c$; $\gtrsim$ 0.05$^d$                \\
W33A                     &  2.0                       & 5.2                              & $\gtrsim$ 2000$^c$; $\gtrsim$ 3$^d$      & $\gtrsim$ 40$^c$; $\gtrsim$ 0.06$^d$                \\
AFGL 2591                &  1.2                       & 2.2                              & $\gtrsim$ 1000$^c$; $\gtrsim$ 2$^d$      & $\gtrsim$ 50$^c$; $\gtrsim$ 0.08$^d$                 \\
\hline
\multicolumn{5}{l}{$^a$ van der Tak, van Dishoek \& Caselli 2000; $^b$ McCall et al. 1999.}\\
\multicolumn{5}{l}{$^b$ Assuming a photon flux in the soft X-ray range of $F_{softX} \gtrsim 2 \times 10^5 $ photons cm$^{-2}$ s$^{-1}$.}\\
\multicolumn{5}{l}{$^c$ $F_{softX} \gtrsim 3 \times 10^2 $ photons cm$^{-2}$ s$^{-1}$.}\\
\end{tabular}
\end{table*}

Methanol is one of the most abundant molecule in interstellar medium and in dense
molecular clouds. Therefore, even despite the reduced production of H$_3^+$ from X-rays
photodissociation process, it is reasonable to expect that at least a fraction of the
detected H$_3^+$ in molecular clouds may be produced from this simple methyl compound
molecule. Considering methanol as an alternative source of H$_3^+$ inside dense
molecular clouds, we derive its reaction scheme by:
\begin{eqnarray}
CH_3OH^+ + h\nu & \stackrel{k_{ph}}{\longrightarrow}   & H_3^+ + COH \quad(\textrm{or } H_3^+ + CO^+) \\
H_3^+ + CO & \stackrel{k_{CO}}{\longrightarrow}   & COX^+ + H_2
\end{eqnarray}
where h$\nu$ is the soft X-ray photons, $k_{ph}$ is the H$_3^+$ photoproduction rate
[s$^{-1}$] for the dissociation of CH$_3$OH and $k_{CO} \sim 2 \times 10^-9$ cm$^3$
s$^{-1}$ is the canonical rate constant for the H$_3^+$ destruction reaction due to CO
molecules. As discussed by McCall et al. (1999), others H$_3^+$ destruction processes
may also occur inside dense molecular clouds, for example, the dissociation due to
electron recombination. However, due to the large abundance of CO in comparison with
free electrons, inside dense molecular clouds, it becomes the dominant destruction route
of H$_3^+$ (Oka 2006). One can ask about the H$_3^+$ photodestruction rate by the same
X-ray field within the cloud, however it certainly must be lower than its UV
photodissociation rate $k_{ph}< 10^{-12}$ s$^{-1}$ (van Dishoeck 1987).

Following the above statements, and assuming the steady-state approximation inside dense
molecular cloud, the rate of change in the amount of H$_3^+$ coming from methanol due to
photodissociation by soft X-rays can be described by:
\begin{equation}\label{eq-imp1}
\frac{d [H_3^+]}{dt} = k_{ph}[CH_3OH] - k_{CO}[CO] = 0
\end{equation}
where [H$_3^+$], [CH$_3$OH] and [CO] is the number density of photoproduced H$_3^+$,
methanol and carbon monoxide inside the molecular cloud.

Assuming a chemical homogeneity into the molecular cloud, Eq.~\ref{eq-imp1} gives the
simple relation:
\begin{equation}\label{eq-imp2}
N^{ph}_{H_3^+} \sim \frac{k_{ph}}{k_{CO}[CO]} N_{CH_3OH}
\end{equation}
where $N^{ph}_{H_3^+}$ is the column density of the photoproduced H$_3^+$ from methanol
dissociation due to soft X-ray (200-310 eV), and $N_{CH_3OH}$ is the column density of
interstellar methyl alcohol. Following Oka (2006), inside dense molecular clouds the CO
number density is about 10-0.1 cm$^{-3}$.

In Table~\ref{tab-N} we present the methanol column density and the total H$_3^+$ column
density, $N_{H_3^+}$, from radioastronomical observations, together with the lower
limits for the H$_3^+$ column density resulting from the photodissociation of methanol
molecules by soft X-rays inside these clouds. The photoproduced H$_3^+$ fraction over
the total observed H$_3^+$ is also showed. We have adopted a typical X-ray luminosity
$L_X \gtrsim 10^{31}$ erg s$^{-1}$ which best fit the observational data for AFGL 2591
(Stäuber et al. 2005) as a good approximation of the radiation field also for the others
dense molecular clouds. Two lower limit for averaged soft X-ray flux were considered,
$F_{softX} \gtrsim 2 \times 10^5 $ and $ \gtrsim 3 \times 10^2 $ photons cm$^{-2}$, as
discussed before. We adopted the averaged CO number density of about 1 cm$^{-3}$.
Assuming a steady state scenario, the highest value for the H$_3^+$ column density from
methanol dissociation by soft X-rays was about $N^{ph}_{H_3^+} \gtrsim 10^{11}$ cm$^2$,
which in the case of AFGL 2591 gives the fraction of the produced H$_3^+$ due to
CH$_3$OH photodissociation, $N^{ph}_{H_3^+}/N_{H_3^+} \>0.05 \%$. For the lower soft
X-ray photon flux the calculated values were about 3 orders of magnitude lower.

Despite the small value of $N^{ph}_{H_3^+}$ compared with the ${H_3^+}$ column density
from radioastronomical observation, this may represent a new source of H$_3^+$ inside
molecular clouds and it is not been considered as yet in interstellar chemistry models.
We cannot firmly assert that this will indeed become necessary to create models which
fully explain radioastronomical observations, but we consider that our experimental
results should be born in mind in those particular circumstances where sources of high
energy excitation of extant CH$_3$X could exist.

Based on our experimental data, we expect that in other methanol rich molecular clouds,
as pointed out recently in 40 galactic center molecular clouds (Requena-Torres et al.
2006), may also have some amount of H$_3^+$ coming from the soft X-ray (or VUV)
photodissociation process of methyl compound molecules.

\section{Conclusions}

As discussed by Ruhl et al. (1990), the formation of double ionized methyl compound
molecules and consequently its dissociation via process like the charge separation and
others, requires excitation energies of the order of 30 eV. Within dense clouds, the
photon energies of this order of magnitude are not normally available because the
absorption by the abundant hydrogen species (and possibly by interstellar grains) in the
other layers. However, the H-atom absorption cross section at wavelengths less than 124
eV ($\sim$ 100 $\AA$) is small enough to allow soft X-rays to penetrate great depths
into these clouds. Therefore, the photochemistry induced by soft X-rays and other
sources of energetic radiation like cosmic rays or stellar wind particles should be
considered within dense clouds.

The present work points out the importance of the dissociation processes promoted by
soft X energy photons on the production of interstellar simply stable polyatomic
molecule, H$_3^+$. The production of H$_3^+$ via the photodissociation of interstellar
methyl compound organic molecules has been studied using photoelectron-photoion
coincidence techniques using soft X-rays photons (200-310 eV). We have shown that the
photodissociation of CH$_3$-X like organic molecules by soft X-rays, leads to the
production of H$_3^+$, among the several other dissociation channels, by:
\begin{equation}
CH_3X + h\nu  \longrightarrow   H_3^+ + CX + e^- \quad (\textrm{or }  H_3^+ + CX^+ +
2e^-)
\end{equation}
where $h\nu$ = soft X-rays and $X$ = OH, NH$_2$ and CN.

We have shown that about 0.7\% and 3.7\% of the single and double photodissociation
channels of methanol, respectively, lead to the H$_3^+$ production. As the methanol is
one of the most abundant molecule in interstellar medium it is reasonable to expect that
at least a fraction of the detected H$_3^+$ could be resultant of this molecule
photodecomposition followed by the rearrangement of the H atoms.

The H$_3^+$ photoproduction cross section due to the dissociation of the studied organic
molecules by photons over the C1s edge (200-310 eV) were about 0.2-1.4 $\times$
10$^{-18}$ cm$^2$.

Assuming a steady state scenario and a typical X-ray luminosity of $L_x \gtrsim 10^{31}$
erg s$^{-1}$ as the case for AFGL 2591 (Stäuber et al. 2005), the fraction of the
produced H$_3^+$ due to CH$_3$OH photodissociation over the observed value, could
reaches up to 0.05\%. Despite the extreme small value, this represent a new and
alternative source of H$_3^+$ inside dense molecular clouds and it is not been
considered as yet in interstellar chemistry models. Moreover, the energetic ionic
products released by dissociation of CH$_3$X molecules, including the H$_3^+$ ion,
become an alternative and efficient route to complex molecular synthesis, since some
ion-molecule reactions do not have an activation barrier and are also very exothermic.

We hope that the H$_3^+$ photoproduction cross section from dissociation of the studied
organic molecules by soft X-rays and the photoproduction rate derived in this work will
give rise to more precise values for some molecular abundances in interstellar clouds
and even in planetary atmosphere models. Better estimative for H$_3^+$ photoproduction
rate depends of more accurate soft X-ray radiation field determinations.

%
\section*{Acknowledgments}
The authors would like to express their gratitude to LNLS staff. This work was supported
by the Brazilian funding agencies FUJB (UFRJ), CAPES, CNPq, and FAPERJ. We are indebted
to the referee Dr. Jonathan Tennyson for the critical reading and essential suggestions
for the manuscript.

%
%

\bsp

\label{lastpage}

\end{document}